\documentclass[11pt, a4paper]{article}
\usepackage{amssymb, geometry, jheppub}                

\title{Fermionic instantons}
\author{N. Houston}
\emailAdd{nhouston@itp.ac.cn}
\affiliation{Kavli Institute for Theoretical Physics China (KITPC) \& Key Laboratory of Theoretical Physics, Institute of Theoretical Physics, Chinese Academy of Sciences, Beijing 100190, P.R. China}
\abstract{We demonstrate the existence of a broad class of non-perturbative fermionic solutions to the Euclidean supergravity equations of motion, which are half-BPS and nonsingular, possess zero action, and obey an (anti)self-duality condition.
These are identified as fermionic instantons associated to the status of the gravitino as the gauge field of local supersymmetry.
By explicitly constructing these configurations from combinations of (anti)self-dual Yang-Mills gauge fields and Killing spinors, we may leverage the ADHM method and generalisations thereof to provide all possible solutions on certain gravitational backgrounds.
As one may expect, these solutions generate and are in turn intrinsically dependent upon spacetime torsion.}

\begin{document}
\maketitle
\section{Introduction}

Across the breadth of theoretical physics and mathematics, one concept leveraged in numerous settings and to numerous ends is that of the instanton.
These are non-singular finite action solutions to Euclidean field equations, which in turn imply an (anti)self-duality condition on the corresponding field strength.

More precisely, in the presence of (anti)self-dual field strength the Yang-Mills equations are satisfied automatically, in the simplest case arising from characteristic field configurations of the form
\begin{align}
	A_a=\frac{1}{g}\frac{x^b\sigma_{ab}}{x^2+\rho^2}\,,
	\label{BPST instanton}
\end{align}
where $g$ is the coupling constant, $\sigma_{ab}$ the self-dual Lorentz generator, and $\rho$ is a scale associated to the size of the instanton.

These solutions approach pure gauge at infinity, and are thereby associated to quantum mechanical tunnelling between topologically distinct vacua indexed by 
\begin{align}
	k\equiv-\frac{1}{4\pi^2}\int_\mathcal{M}\, \mathrm{Tr}\left(F\wedge F\right)\,,\quad
	F\equiv dA+A\wedge A\,.
	\label{Pontryagin number}
\end{align}
As canonical examples of non-perturbative phenomena, they are also invisible in perturbation theory by virtue of their inverse dependence on $g$.

Originally elucidated more than forty years ago \cite{Belavin:1975fg}, these field configurations are now known to be a characteristic feature of the QCD vacuum, thereby responsible for many aspects of the resulting phenomenology \cite{Schafer:1996wv}, and have found further application across a wide swathe of topics in theoretical physics \cite{Bianchi:2007ft}.
In addition to their physical utility, instantons have also found an array of applications in purely mathematical contexts \cite{Freed:1984xe}.

Furthermore, we may note that these concepts can be usefully imported into the gravitational context, given that the Einstein field equations are also automatically satisfied in the presence of (anti)self-dual Riemann curvature.
This leads to the notion of gravitational instantons; finite action asymptotically locally Euclidean solutions of the Einstein equations, indexed respectively by the Euler number and Hirzebruch signature
\begin{align}
	\chi\equiv\frac{1}{16\pi^2}\int_\mathcal{M} \,R_{ab}\wedge*R^{ab}\,,\quad
	\tau\equiv-\frac{1}{24\pi^2}\int_\mathcal{M} \,\mathrm{Tr}\left(R_{ab}\wedge R^{bc}\right) \,,
	\label{topological invariants}
\end{align}
where $*$ is the Hodge dual. 
Originating from the same era and considerations as the analogous gauge theory solutions \cite{Eguchi:1978gw}, known examples of gravitational instantons also find wide application across a variety of contexts.

The line of reasoning connecting both of these scenarios is the status of the underlying degrees of freedom as gauge fields, or in more mathematical terms, connections on principal bundles. 
In the former case $A$ is of course the familiar gauge field of Yang-Mills theory, and in the latter we have the spin connection $\omega^{ab}$, which can be thought of as arising from the gauging of Lorentz symmetry into the full diffeomorphism group.

As is well established, given supersymmetry transformations parametrised by a spinor $\epsilon$, promoting this field to a function of local coordinates also necessarily mandates a gauge field, the spin 3/2 gravitino $\psi$.
This may be accounted for via the observation that if supersymmetry is made local we then expect a new transformation $\delta\left(\dots\right)\sim \partial_\mu\epsilon\left(x\right)$, implying the existence of a previously absent field carrying both a vector and a spinor index.

This perspective suggests the existence of a previously unexplored third class of instanton, complementary to the gauge and gravitational phenomena outlined above.
This being the case, the considerations which have been so useful elsewhere may bear yet more fruit when transplanted into this novel context.

Indeed, in the following we will establish that:
\begin{enumerate}
	\item Supergravity configurations $\left(e^a, \psi\right)$ for which the torsion-free Riemann curvature obeys a duality condition and appropriate boundary conditions, whilst the gravitino generates torsion and is constructed via derivatives of torsion-free Killing spinors, automatically satisfy the associated equations of motion with zero action and are half-BPS.
	\item By constructing the gravitino in this fashion we may import many of the notions and techniques of Yang-Mills instanton calculus, such that the task of finding explicit field configurations satisfying these conditions can be reduced in certain backgrounds to a problem for which the solution is already known.
	\item This results in characteristic solutions which are schematically of the form
		\begin{align}
			\psi_\mu=\frac{1}{\kappa}A_\mu\epsilon\,,\nonumber
		\end{align}
	where $A_\mu$ is an (anti)self-dual Yang-Mills gauge field, $\epsilon$ is a Killing spinor of the corresponding torsion-free background, and $\kappa^2=8\pi G$ is the usual gravitational coupling. 
	We identify these as fermionic instantons associated to the status of the gravitino as the gauge field of local supersymmetry.
\end{enumerate}

Although there exists a large body of literature on exact supergravity solutions, including a number of comprehensive results \cite{Gibbons:1982fy, Tod:1983pm, FigueroaO'Farrill:2002ft, Gauntlett:2002fz, Gauntlett:2002nw, Gutowski:2003rg, Bellorin:2005zc, Cacciatori:2007vn, Bellorin:2007yp, Gran:2008vx}, these solutions are overwhelmingly of the `bosonic' type, where fermionic backgrounds are trivial.
A significantly smaller fraction of this research effort involves non-trivial solutions for the gravitino field \cite{Urrutia:1981sk, Hull:1983aw, Hull:1984vh, Aichelburg:1986wv}, albeit also often neglecting the non-linear, and in this context essential, effect of gravitino-induced spacetime torsion.
To the best of our knowledge the solutions presented herein, their method of construction and indeed the notion of fermionic instantons, are novel.

\section{Preliminaries}

Before proceeding further we will establish some notation and conventions for what follows.
As one may imagine, there exist a number of non-trivial aspects to solving the fully non-linear supergravity equations of motion.
These may be ameliorated to some extent by framing the problem appropriately.

\subsection{Duality structure}
We may firstly note that in Euclidean signature
the Hodge star satisfies the condition $*^2=1$, implying a canonical decomposition on the bundle of two-forms
\begin{align}
	\Lambda^2=\Lambda_+^2\oplus\Lambda_-^2\,, 
	\label{decomposition}
\end{align}
where $\Lambda^2_\pm$ are the $\pm1$, or equivalently self and antiself-dual, eigenspaces of $*$.
For concreteness we will specialise in the following to self-dual instanton configurations only.

Whilst independently useful, this decomposition also implies the group isomorphism 
\begin{align}
	\mathrm{Spin}\left(4\right)\simeq SU\left(2\right)_+\times SU\left(2\right)_-\,,
\end{align}
in terms of which the underlying problem may be usefully expressed.
The corresponding self and antiself-dual generators are
\begin{align}
	\sigma_{ab}\equiv\frac{1}{4}\left(\sigma_a\overline\sigma_b-\sigma_b\overline\sigma_a\right)\,,\quad
	\overline\sigma_{ab}\equiv\frac{1}{4}\left(\overline\sigma_a\sigma_b-\overline\sigma_b\sigma_a\right)\,,
\end{align}
where $\left(\sigma_a\right)_{A\dot A}=\left(i\vec\tau,1_{2\times2}\right)$ and $\left(\bar\sigma_a\right)^{\dot AA}=\left(-i\vec\tau,1_{2\times2}\right)$ are the Euclidean sigma matrices, with $\vec\tau$ the ordinary Pauli matrices, and $A, \dot A=\{1,2\}$ are respectively the spinor indices of $SU\left(2\right)_+$ and $SU\left(2\right)_-$.
We will use lowercase roman characters $\{a,b,\dots\}$ for tangent space indices and greek characters $\{\alpha,\beta,\dots\}$ for spacetime indices, with three-vector indices exclusively represented via $\{i,j,k\}$.
Following the conventions of \cite{Dorey:2002ik}, our Euclidean gamma matrices are then
\begin{align}
	\gamma^a=
	\begin{pmatrix}
		0 & -i\sigma^a\\
		i\overline\sigma^a & 0\\
	\end{pmatrix}\,,
	\label{gamma matrix convention}
\end{align}
such that we have the useful chiral-duality identity 
\footnote{\,We (anti)symmetrise always with unit weight, such that e.g. $\gamma^{[mn]}=\frac{1}{2}\left(\gamma^{mn}-\gamma^{nm}\right)$.}
\begin{align}
	\gamma^{ab}\equiv
	\gamma^{[a}\gamma^{b]}
	=\frac{1}{2}\gamma^5\epsilon^{abcd}\gamma_{cd}\,.
	\label{chiral-duality identity}
\end{align}

\subsection{Euclidean fermions}
This structure is further reflected in the standard two-component notation for Dirac spinors, where fields in the $\left(\frac{1}{2},0\right)\oplus\left(0,\frac{1}{2}\right)$ representation are written
\begin{align}	
	\psi=
	\begin{pmatrix}
		\chi_{A}\vspace{2pt}\\
		\bar{\xi}^{\dot A}
	\end{pmatrix}=
	\begin{pmatrix}
		\psi_+\vspace{2pt}\\
		\psi_-
	\end{pmatrix}\,,\quad
	\chi\phi\equiv\chi^A\phi_A\,,\quad
	\bar\xi\bar\phi\equiv\bar\xi_{\dot A}\bar\phi^{\dot A}\,,
	\label{fermion components}
\end{align}
with $\psi_+$ and $\psi_-$ independent, and so labelled in accordance with the usual projection operators $P_\pm=\frac{1}{2}\left(1\pm\gamma^5\right)$.
Although this fits neatly, there does exist however a well-known problem in this context; that of fermions in Euclidean signature.

It is first of all notable that there are no Majorana representations in Euclidean signature since $SU\left(2\right)_+$ and $SU\left(2\right)_-$ do not mix under conjugation.
This precludes the Majorana reality condition which would otherwise relate $\psi_+$ and $\psi_-$, which is of course problematic for theories exhibiting $\mathcal{N}=1$ supersymmetry, reliant as they are upon such representations.
As such, we will necessarily promote our gravitino from a Majorana field to the Dirac gravitino typically encountered in $\mathcal{N}=2$ supergravity, with the hope that whatever results we derive may still be of relevance to the $\mathcal{N}=1$ case
\footnote{\,At any rate, the solutions we ultimately find will only preserve half of the underlying $\mathcal{N}=2$ supersymmetry as a consequence of self-duality.}.

Unfortunately, there are yet further issues in that the Lorentzian inner product, $\overline\psi\psi\equiv\psi^\dagger\gamma^0\psi$, is not preserved in Euclidean signature, since $\gamma^0$ has the effect of mixing dotted and undotted spinors.
Therefore, we should apparently instead utilise the $\mathrm{Spin}\left(4\right)$ invariant $\psi^\dagger\psi=\psi_+^\dagger\psi_++\psi_-^\dagger\psi_-$ which comes at the price, given the absence of $\gamma^0$, of the loss of a definite Hermiticity property of any resultant action.

As discussed in \cite{Dorey:2002ik}, there exist a number of workarounds and perspectives on this issue. 
The most conservative viewpoint, which we will espouse in the following, is that our primary concern is simply the computation of Minkowski space Green's functions.
The standard Wick rotation into Euclidean signature offers a convenient and well-defined method of calculation, from which these can be extracted.
That fermionic actions cease to be real-valued after Wick rotation is then something of an irrelevance, in that we can regardless arrive at the correct Minkowski signature result.

It does however result in a further non-triviality in that, under the standard Euclideanisation prescription \cite{Osterwalder:1973dx}, this implies that the number of fermionic degrees of freedom, Dirac or otherwise, must be effectively doubled.
In practice this amounts to replacing $\psi^\dagger$ with a $\psi$-independent field $\chi^\dagger$, although the number of Grassman integration variables ultimately remains unchanged.

However, it should be noted that there also exist a variety of other techniques for the Euclideanisation of fermions \cite{Fubini:1972mf, Zumino:1977yh, Nicolai:1978vc,Mehta:1986mi,vanNieuwenhuizen:1996tv}, some of which do not involve any kind of doubling, and some of which do preserve Hermiticity. 
Furthermore, at present there seems to be no clear consensus on a `best' approach, resulting in a rather unclear situation overall.

As such it seems best to take an agnostic perspective on the issue, and eschew arguments entirely which rely on a choice of inner product.
For notational familiarity we will still write barred spinors such as $\overline\psi$, but without assuming any precise relationship to $\psi$.

\subsection{Cartan formalism}
On the geometrical side we will also require some standard results from the Cartan formalism; given a set of frame one-forms $e^a=e^a{}_\mu dx^\mu$, the first Cartan structure equation provides the torsion
\begin{align}
	T^a\equiv de^a+\omega^{ab}\wedge e_b\,,
	\label{first Cartan}
\end{align}
in terms of a spin connection $\omega^{ab}$,
which then yields the Riemann curvature via
\begin{align}
	R^{ab}\equiv d\omega^{ab}+\omega^a{}_c\wedge\omega^{cb}\,.
\end{align}
In practice it will be expedient to decompose the spin connection into
\begin{align}
	\omega^{ab}=\omega_\circ^{ab}+K^{ab}\,,\quad
	de^a+\omega_\circ^{ab}\wedge e^b=0\,,\quad
	T^a=K^{ab}\wedge e^b\,,
	\label{decomposed first Cartan}
\end{align}
where $\omega_\circ^{ab}$ is the unique torsion-free spin connection, and $K^{ab}$ is the contortion one-form.

In this notation the supergravity Lagrangian is
\begin{align}
	\mathcal{L}=\frac{1}{2}\epsilon_{abcd} e^a\wedge e^b\wedge R^{cd}-\overline\psi \wedge \gamma^5\gamma\wedge D \psi\,,\quad
	D\psi\equiv d\psi +\frac{1}{4}\omega^{ab}\gamma_{ab}\wedge\psi\,,
	\label{supergravity Lagrangian}
\end{align}
where we have the following definitions
\begin{align}
	\gamma\equiv \gamma_ae^a\,,\quad
	\gamma^5\equiv
	\begin{pmatrix}
		1 & 0\\
		0 & -1\\
	\end{pmatrix}\,,\quad
	\psi\equiv \psi_a e^a=\psi_\mu dx^\mu\,,\quad
	\label{definitions}
\end{align}
and invariance under the transformations
\begin{align}
	\delta \psi= D\chi\,,\quad
	\delta e^a =\frac{1}{2}\left(\overline\chi\gamma^a\psi-\overline\psi\gamma^a\chi\right)\,.
	\label{supersymmetry variations}
\end{align}
To streamline presentation we are evidently neglecting the $U\left(1\right)$ graviphoton field of $\mathcal{N}=2$ supergravity, although it may of course be straightforwardly incorporated in what follows.

Since it will be of relevance later on, we also note that the Rarita-Schwinger prefactor here is the correct choice for a Dirac field.
A Majorana gravitino would require the more conventional factor of 1/2 to correct for overcounting arising from self-conjugacy.

\section{Fermionic instantons}

Having established sufficient formalism, we may now search for the instanton solutions which the arguments of the introduction suggest should exist.

\subsection{Chiral-duality symmetry}
We firstly recall that since both the Einstein and Yang-Mills equations are automatically satisfied given a duality condition on the field strength, it is sensible to pursue a similar result for the Rarita-Schwinger equations,
\begin{align}
	\gamma\wedge D\psi
	=\frac{1}{2}\gamma_\alpha\psi_{\mu\nu} \,e^\alpha\wedge e^\mu\wedge e^\nu
	=0\,,\quad
	\psi_{\mu\nu}\equiv D_{\mu}\psi_{\nu}-D_{\nu}\psi_{\mu}\,.
	\label{Rarita-Schwinger equations 1}
\end{align} 
By contracting the components with $\epsilon^{\alpha\beta\mu\nu}$ we can arrive at two equivalent relations,
\begin{align}
	\gamma^\mu\psi_{\mu\nu}=0\,,\quad
	\gamma^\mu*\psi_{\mu\nu}=0\,,\quad
	{}*\psi_{\alpha\beta}\equiv\frac{1}{2}\epsilon^{\alpha\beta\mu\nu}\psi_{\mu\nu}\,,
	\label{Rarita-Schwinger equations 2}
\end{align}
which are indeed suggestive of some underlying duality structure.
Use of the chiral-duality identity \eqref{chiral-duality identity} and $\gamma^\mu\gamma^\nu=\gamma^{\mu\nu}+\delta^{\mu\nu}$ further yields
\begin{align}	
	\gamma_\alpha\gamma^\beta\psi_{\beta\mu}
	=\psi_{\alpha\mu}+\gamma^5\left(\gamma^\mu\gamma^\nu *\psi_{\alpha\nu}-{}*\psi_{\alpha\mu}\right)\,,
\end{align}
indicating that the equations in \eqref{Rarita-Schwinger equations 2} imply, and are in turn interrelated by, the chiral-duality symmetry
\footnote{\,Care should be exercised in comparison with the literature in that \eqref{chiral-duality identity} is signature-dependent, such that $\psi_{\mu\nu}=-\gamma^5*\psi_{\mu\nu}$ in the Lorentzian case  \cite{Deser:1977ur}.}
\begin{align}
	\psi_{\mu\nu}=\gamma^5*\psi_{\mu\nu}\,.
	\label{chiral-duality symmetry}
\end{align}
In form notation this is equivalent to
\begin{align}
	\begin{pmatrix}
	D\psi_+\\
	D\psi_-
	\end{pmatrix}
	=
	\begin{pmatrix}
	+*D\psi_+\\
	-*D\psi_-
	\end{pmatrix}\,,
	\label{chiral-duality}
\end{align}
which we may recognise as self and anti-self duality in the positive and negative chirality projections, respectively. 
This then indicates that in fact every solution of the Rarita-Schwinger equations obeys a duality condition, but also that this chiral-duality is not sufficient in and of itself to ensure solution.

Building upon the reasoning surrounding instanton solutions in general relativity and Yang-Mills theory, we therefore require some further condition to delineate those field configurations which are `special' in some sense; in that they possess non-trivial topological character and automatically solve the equations of motion.

\subsection{Geometric \& topological considerations} 
To find such a condition we may note that for any spinor $\epsilon$ the ansatz $\psi=D\epsilon$ must always be a valid solution, since $\delta\psi=D\epsilon$ under supersymmetry.
Clearly some subset of these solutions are trivial, in that they are merely gauge transformations of $\psi=0$.
To delineate these possibilities, we may specialise for concreteness to spacetimes which locally approximate flat spacetime at infinity, and examine their asymptotic behaviour.
\begin{enumerate}
	\item If $\epsilon$ is first of all an increasing function at infinity, then $\psi$ will not be normalisable and hence not a valid solution for our purposes.
	\item If $\epsilon$ vanishes at infinity, then the associated transformation will be asymptotic to the identity. $\psi$ will therefore be pure gauge and hence may be removed by gauge fixing.
	\item If $\epsilon$ is asymptotic to a constant value at infinity, it is non-normalisable and as such inadmissible as a gauge parameter.
	Since $D\epsilon$ will nonetheless vanish at infinity, $\psi$ itself can in contrast be normalisable.
	This is the subset of solutions we will consider.
\end{enumerate}
We need only consider asymptotic behaviour since the difference of any two $\psi$ in the same asymptotic class must vanish at infinity, and so can be removed via gauge transformation.

As observed in \cite{Witten:1981nf}, it is only the latter class which incorporate zero modes associated to supersymmetry transformations.
This may be considered a foreshadowing of the zero modes we expect any instanton solution to possess as a result of collective coordinate invariance.

It is first of all evident that we expect instantons to minimise the energy within a given homotopy class, and in so doing saturate a Bogomol'nyi bound.
In general relativity the corresponding relation is
\begin{align}
	M\geq\sqrt{Q^2+P^2}\,,
\end{align}
where $M$ is the ADM mass of the spacetime, and $Q$ and $P$ are the electric and magnetic charges respectively.

As demonstrated in \cite{Gibbons:1982fy}, saturation of this bound implies the existence of solutions to the equation $D\epsilon=0$.
These Killing, or equivalently supercovariantly constant, spinors are of particular importance in supergravity and indeed differential geometry in general. 
Their relevance to our picture is as follows.

Since the metric must approximate locally flat spacetime at infinity, whatever non-triviality is present must be suppressed by powers of $r$.
Given that the connection components are defined via derivatives of the metric, their contribution at infinity must therefore be at most $\mathcal{O}\left(1/r^2\right)$.
Evidently the Killing spinor equation at infinity is then $d\epsilon=0$, which implies that any non-trivial solution $\epsilon$ provides a characteristic representative of the class of spinors which are constant at infinity.

Relative to our original delineation, this appears to be paradoxical.
If $\epsilon$ vanishes at infinity, then $\psi$ is pure gauge and therefore trivial.
If however $\epsilon$ is constant at infinity, then it is related by gauge transformation to a Killing spinor, and $\psi=D\epsilon$ again vanishes identically and is unsuitable for our purposes.

The resolution to this quandary lies in the fact that our argument is valid for general relativity only, and so neglects spacetime torsion. 
In supergravity, as with any theory of gravity coupled to fermions, the possibility of this torsion must be accounted for.
It is furthermore inescapable if, as in the present setting, one seeks non-trivial background configurations for the fermion fields.

If, as before, $\epsilon$ is Killing relative to the torsion-free background we then have
\begin{align}
	\psi=D\epsilon
	=\left(d+\frac{1}{4}\gamma_{ab}\left(\omega^{ab}_\circ+K^{ab}\right)\right)\epsilon
	=\frac{1}{4}\gamma_{ab}K^{ab}\epsilon\,,
	\label{Killing-modulo-torsion}
\end{align}
where we have decomposed the full spin connection into the unique torsion-free part $\omega_\circ^{ab}$, and the contortion one-form $K^{ab}$, which ultimately will be sourced by the gravitino.
The apparent overlap between the pure gauge and non-gauge asymptotic classes is then seen to be a consequence of having implicitly selected the zero instanton sector, when $\gamma_{ab}K^{ab}$ vanishes.

Despite no longer satisfying $D\epsilon=0$, spinors such as $\epsilon$ are nonetheless sensitive to the underlying geometric information in that their bilinears still form Killing vectors via
\begin{align}
	\xi_\nu=\overline\epsilon\, e^a{}_\nu\gamma^a\,\epsilon\,,\quad
	\nabla_\mu\xi_\nu+\nabla_\nu\xi_\mu=0\,.
	\label{Killing equation}
\end{align}
This follows from the fact that the torsion resides in the antisymmetric part of the affine connection, which is removed by the symmetry of Killing's equation.

\subsection{Connecting connections}

Given the somewhat slippery nature of spacetime torsion, it is not necessarily clear how we should interpret \eqref{Killing-modulo-torsion} in relation to the more familiar Yang-Mills instanton solutions.
We may however make a connection explicit in the following way.

By virtue of \eqref{chiral-duality identity} the positive chirality gravitino satisfies
\begin{align}
	\gamma^5\psi_+
	=\gamma_{cd}K^{cd}\epsilon_+
	=\frac{1}{2}\epsilon^{abcd}\gamma_{cd}K_{ab}\epsilon_+\,,
\end{align}
which is eminently suggestive of a self-duality condition on the contortion one form,
\begin{align}
	K^{cd}=\frac{1}{2}\epsilon^{abcd}K_{ab}\,.
	\label{self-dual contortion 1}
\end{align}
Taking some inspiration from the approach of \cite{Charap:1977ww}, this may be engineered by interpreting the contortion as a gauge field via
\begin{align}
	K^{ab}=A^i\eta^{iab}\,,
	\label{contortion gauge field}
\end{align}
where $A^i$ is a one-form carrying a single $\mathfrak{su\left(2\right)}$ index
and $\eta^{iab}$ is the self-dual 't Hooft symbol. 
We then have
\begin{align}
	\frac{1}{4}K^{ab}\gamma_{ab}
	=\frac{1}{2}K^{ab}\sigma_{ab}
	=A^i\sigma_i
	\equiv A\,,
	\label{spin connection embedding}
\end{align}
where we have made use of the identity $\sigma_{ab}=\frac{1}{2}\eta_{jab}\sigma^j$, and
\begin{align}
	K^{ab}\gamma_{ab}
	=\frac{1}{2}\left(K^{ab}\gamma_{ab}+\frac{1}{2}\epsilon^{abcd}K_{cd}\gamma_{ab}\right)
	=\frac{1}{2}K^{ab}\gamma_{ab}\left(1+\gamma_5\right)\,.
	\label{self-dual contortion 2}
\end{align}
Equation \eqref{Killing-modulo-torsion} then implies 
\begin{align}
	\psi_+=A\epsilon_+\,,\quad
	D\psi_+
	=d\left(A\epsilon_+\right)+\left(\frac{1}{4}\gamma_{ab}\omega_\circ^{ab}+A\right)\wedge A\epsilon_+
	=\left(dA+A\wedge A\right)\epsilon_+\,.
	\label{Dpsi F relation}
\end{align}
Compatibility with the underlying chiral-duality of the Rarita-Schwinger equations evidently enforces self-duality of $F\equiv dA+A\wedge A$, as a consequence of \eqref{chiral-duality}.
This is however unproblematic as we already know how to construct self-dual gauge fields, and hence the corresponding gravitino configurations, via the ADHM method \cite{Atiyah:1978ri} and generalisations thereof.

This established, we postpone further exploration of this perspective until later on; before proceeding there are a number of important aspects we must firstly take account of.

\subsection{Field strength}

For any $\epsilon$, the associated gravitino field strength is
\begin{align}
	D\psi
	=D^2\epsilon
	=\frac{1}{4}R^{ab}\gamma_{ab}\epsilon\,,
	\label{integrable Dpsi 1}
\end{align}
one side of which is just the standard integrability condition used in the characterisation of supersymmetric backgrounds
\footnote{\, Using the coordinate covariant derivative $\nabla$ instead yields extra terms proportional to $T^a$ \cite{Freedman:2012zz}.}.
Since we require the existence of solutions to $D\epsilon_+=0$ in the torsion-free limit, any torsion-free Riemann curvature must satisfy $R_\circ^{ab}\gamma_{ab}\epsilon_+=0$.
This requirement is restrictive, but may be engineered in the non-trivial case if the curvature satisfies an antiself-duality condition, which implies
\begin{align}
	R_\circ^{ab}\gamma_{ab}
	=R_\circ^{ab}\gamma_{ab}P_-\,,
	\label{self-dual Riemann}
\end{align}
as follows from the same argument used in \eqref{self-dual contortion 2}.
A useful corollary \cite{Atiyah:1978wi} of this result is that given antiself-dual Riemann curvature, there always exists a frame in which the spin connection satisfies the analogous relations
\begin{align}
	\omega^{ab}_\circ=-\frac{1}{2}\epsilon^{abcd}\omega^{cd}_\circ\,,\quad
	\omega^{ab}_\circ\gamma_{ab}
	=\omega^{ab}_\circ\gamma_{ab}P_-\,.
	\label{antiself-dual spin connection}
\end{align}
The condition $D\epsilon_+=d\epsilon_+=0$ is now trivial, since we may always find constant spinors.
Returning to the torsionful context, we may note an interesting consequence of the contrasting duality structures of $K^{ab}$ and $\omega_\circ^{ab}$.
In light of our underlying decomposition into self and antiself-dual two-forms, encoded in \eqref{decomposition}, the torsionful Riemann curvature should similarly factorise into
\begin{align}
	R^{ab}=R_\circ^{ab}+F^i\eta^{iab}\,.
	\label{Riemann decomposition}
\end{align}
In practice we may verify by inserting \eqref{self-dual contortion 1} and \eqref{antiself-dual spin connection} and contracting in the cross terms
\begin{align}
	\omega_\circ^{ae}\wedge K^{eb}-\left(a\leftrightarrow b\right)
	=\frac{1}{4}\delta^{[a}{}_{[b}\delta^c{}_f\delta^{e]}{}_{g]}\omega_\circ^{ce}\wedge K^{fg}-\left(a\leftrightarrow b\right)
	=\omega_\circ^{be}\wedge K^{ea}-\left(a\leftrightarrow b\right)\,,
\end{align}
which obviously vanishes.
This then allows the convenience of separating the torsionful and torsion-free aspects of the analysis.

Returning to the gravitino, we already know that $\psi_+=D\epsilon_+=A\epsilon_+$ and so it simply remains to construct $\psi_-$.
Since the arguments of the previous section suggest that the solutions we seek should be trivial in the torsion-free limit, which in any case has no effect on $\psi_-$ thanks to \eqref{self-dual contortion 2}, it is evident that we require
\begin{align}
	D\psi_-=d\psi_-+\frac{1}{4}\gamma_{ab}\omega_\circ^{ab}\wedge\psi_-=0\,,
\end{align}
and likewise for $D\overline\psi_-$.
In either case we may construct an appropriate solution incorporating again a constant anticommuting $\epsilon_+$, via
\begin{align}
	\psi_{-}^{\dot A}
	=2ie^a\left(\overline\sigma_a\right)^{\dot A A}\left(\epsilon_+\right)_A
	=2e^a\gamma_a\epsilon_+\,,
\end{align}
where we have reinstated spinor indices for clarity.
Since negative chirality spinors such as $\gamma^a\epsilon_+$ are insensitive to the spacetime torsion $T^a\equiv De^a$ here, $D\psi_-$ should evidently vanish.
In practice $\overline\psi_-$ will actually be more important to us, the same argument gives
\begin{align}
	\overline\psi_{-\dot A}
	=2i\epsilon_+^A\left(\sigma_a\right)_{A\dot A}e^a
	=-2\epsilon_+\gamma_ae^a\,.
	\label{overline psi}
\end{align}

In light of our preliminary comments on Euclidean spinors it isn't entirely clear if we require some fixed relationship between barred and unbarred spinors here, although thankfully this isn't necessarily problematic. 
Either we may independently set $\overline\psi_+$ and $\psi_-$ to zero, or, if some conventional mapping exists their contribution will in any case mirror that of $\psi_+$ and $\overline\psi_-$. 
As such we restrict attention to $\psi_+$ and $\overline\psi_-$ in the following, this being the minimal set of spinors we require.

\subsection{Action \& equations of motion} 
Given that $\psi_+=D\epsilon_+$, consistency of the theory should guarantee solution of the equations of motion, since the variation of the action under local supersymmetry is now proportional to the corresponding Euler-Lagrange equations. 
We may nonetheless verify.

Varying the Rarita-Schwinger Lagrangian with respect to the gravitino, \eqref{Dpsi F relation} provides
\begin{align}
	\delta\overline\psi_-\wedge\gamma^5\gamma\wedge D\psi_+
	=-\delta\overline\psi_-\wedge\gamma\wedge \left(dA+A\wedge A\right)\epsilon_+\,.
\end{align}
The latter term may be rewritten, noting \eqref{self-dual contortion 2}, via
\begin{align}
	\frac{1}{4}\gamma^b\gamma_{cd}\,K^{cd}\epsilon_+
	=\frac{1}{4}\left(\gamma^{b}{}_{cd}+2\delta^{b}{}_{[c}\gamma_{d]}\right)K^{cd}\epsilon_+
	=\gamma_{d}\,K^{bd}\epsilon_+
	=P_-\gamma_{d}\,K^{bd}\epsilon_+\,,
	\label{three gamma identity}
\end{align}
where $\gamma^{abc}\equiv\gamma^{[a}\gamma^b\gamma^{c]}=\epsilon^{abcd}\gamma_d\gamma^5$, giving
\begin{align}
	\frac{1}{4}\gamma\wedge A\wedge A\epsilon_+
	=\gamma^a K^{ab}\wedge e^b\wedge\psi_+
	=\gamma^aT^a\wedge\psi_+\,.
\end{align}	
This vanishes when the torsion constraint is satisfied thanks to the Fierz identity
\begin{align}
	\gamma^a\left(\overline\psi\wedge\gamma^a\psi\right)\wedge\psi
	=0\,,
\end{align}
which unsurprisingly also appears in proofs of local supersymmetry invariance \cite{Freedman:2012zz}.
Also, 
\begin{align}
	d\left(\overline\psi_-\wedge\gamma\right)
	=2d\left(\epsilon_+\gamma^a\gamma^b e^b\wedge e^a\right)
	=-4\epsilon_+\gamma^a\gamma^b\omega_\circ^{bc}\wedge e^c\wedge e^a
	=0\,,
\end{align}
since, by the analogue of \eqref{three gamma identity}, $\gamma^b\omega_\circ^{bc}=P_+\gamma^b\omega_\circ^{bc}$.
Given that $d\delta\overline\psi_-=\delta d\overline\psi_-$, the remaining $dA$ term may therefore be rewritten as a total derivative.
Solution of the Rarita-Schwinger equations is then assured.

Since the torsion-free curvature obeys a duality condition, it follows that the associated equations of motion are satisfied with zero action, by virtue of the first Bianchi identity
\begin{align}
	R^{ab}_\circ\wedge e^b\
	=0\,.
\end{align}
As such, we need consider only the torsionful piece of the Einstein-Hilbert action.
Noting that the Rarita-Schwinger action vanishes on shell, we may expand
\begin{align}
	\overline\psi_-\wedge\gamma^5\gamma\wedge D\psi_+
	=\frac{1}{2}\epsilon_+\gamma_a\gamma_b\gamma_{cd}\,e^a\wedge e^b\wedge F^i\eta^{icd}\epsilon_+
	=2\epsilon_+\gamma_a\gamma_c\,e^a\wedge e^b\wedge F^i\eta^{ibc}\epsilon_+\,,
\end{align}
where we have used the $\eta^{icd}$ analogue of \eqref{three gamma identity}.
The Fierz identity 
\begin{align}
	\chi_1\sigma^{ab}\chi_2=-\chi_2\sigma^{ab}\chi_1\,,
	\label{Fierz identity}
\end{align}
then removes the antisymmetric piece of $\gamma_a\gamma_c$, leaving
\begin{align}
	2e^a\wedge  e^b\wedge F^i\eta^{iba}\epsilon_+\,\epsilon_+
	=-\epsilon_{abcd}\, e^a\wedge e^b\wedge F^i\eta^{icd}\epsilon_+\,\epsilon_+\,,
\end{align}
where similarity with the torsionful piece of the Einstein-Hilbert Lagrangian is self-evident.
Since we may assume the normalisation $\left(\epsilon_+\right)^A\,\left(\epsilon_+\right)_A=1$ without loss of generality, the on-shell Einstein-Hilbert action evidently vanishes too.
Repeating these steps from a slightly different starting point also yields
\begin{align}
	\overline\psi_-\wedge\gamma^5\gamma_b\delta e^b\wedge D\psi_+
	=\epsilon_{abcd}\, \delta e^a\wedge e^b\wedge F^i\eta^{icd}\,,
	\label{Einstein equations}
\end{align}
where we may recognise the latter term from variation of the Einstein-Hilbert Lagrangian.
Given the relative sign in \eqref{supergravity Lagrangian}, this then implies that the Einstein equations are also automatically satisfied.

A similar argument can be made to demonstrate invariance under $\delta\omega^{ab}$, and hence solution of the algebraic torsion constraint, but it is straightforward to simply note that 
\begin{align}
	\frac{1}{2}\overline\psi_-\wedge\gamma^b\psi_+
	=-\frac{1}{4}\epsilon_+\gamma_a e^a\gamma^b\gamma^{cd}\wedge K^{cd}\epsilon_+
	=K^{ba}\wedge e^a
	=T^b\,,
	\label{algebraic torsion constraint}
\end{align}
as required, where we have used \eqref{three gamma identity} and \eqref{Fierz identity}
\footnote{\,As the Einstein-Hilbert action yields a factor of two when varying $e^a$ or $\omega^{ab}$, whilst the Rarita-Schwinger action does not, we can be confident the correct $\overline\psi_-$ prefactor for \eqref{Einstein equations} is the same as for \eqref{algebraic torsion constraint}.}.

\subsection{Supersymmetry invariance} 

In addition to satisfying the equations of motion, we furthermore require invariance under transformations of the form of \eqref{supersymmetry variations}.
Solutions which saturate a Bogomol'nyi bound are of course BPS and so should preserve only some fraction of the underlying supersymmetry, which in practice amounts to non-conservation of some of the supercharges $Q_\pm$.

In the context of `bosonic' supergravity solutions the background value of the gravitino is set to zero, so $\delta e^a$ is also zero and we need only demonstrate invariance under $\delta\psi$ to determine the residual global supersymmetry algebra of any solution.
Such an approach also evades the complications of spacetime torsion in that the background value of the contortion tensor will be zero.
In the present context, since we are instead seeking topologically non-trivial configurations for the gravitino field, no such simplification is possible.

There is however an obvious generalisation from the antiself-dual bosonic case, where we need only demonstrate that $\delta\psi_+$ is a symmetry of the theory, to the requirement of invariance under both constant shifts generated by $Q_+$,
\begin{align}
	\delta\psi_+
	=D\chi_+,\quad
	\delta e^a
	=-\frac{1}{2}\overline\psi_-\gamma^a\chi_+\,.
\end{align}
Since $\chi_+$ is constant it must be proportional to $\epsilon_+$, which then implies 
\begin{align}
	\delta\psi_+
	\sim D\epsilon_+
	=\psi_+,\quad
	\delta e^a
	\sim-\frac{1}{2}\overline\psi_-\gamma^a\epsilon_+
	=\epsilon_+\gamma^b\gamma^a\epsilon_+e^b
	=e^a\,,
\end{align}
where we have used the Fierz identity \eqref{Fierz identity}.
Evidently any solution is simply rescaled by some constant value and, as the constant of proportionality is the same for both fields, varying the total action simply rescales it by a fixed amount.
Invariance is nonetheless assured as the action vanishes regardless.

In the context of the full $\mathcal{N}=2$ supergravity, invariance of the $U\left(1\right)$ graviphoton field $V$ is also required. 
It is however straightforward to note via \eqref{Fierz identity} that under the corresponding $Q_+$ transformation
\begin{align}
	\delta V
	=\frac{1}{2}\chi_+\,\psi_+
	\sim\epsilon_+\gamma^{ab}\epsilon_+K^{ab}
	=0\,.
\end{align}
As such, invariance under half of the underlying supersymmetry is still preserved.

\subsection{Explicit solutions}

These points established, we may now return to constructing solutions and their properties.
To fully isolate the fermionic degrees of freedom, we will specialise presently to the case of zero spacetime curvature.
In this case $\omega_\circ^{ab}$ vanishes, and the Riemann tensor becomes
\begin{align}
	R^{ab}=F^i\eta^{iab}\,.
	\label{F_K}
\end{align}
Whilst this background is certainly flat, we of course also require it to differ topologically from Euclidean spacetime
\footnote{\,Non-trivial topology needn't be an obstruction to the applicability of conventional instanton techniques, in that we may recognise a parallel between this situation and that of instanton calculus in singular gauge.
More specifically, on a formal level the use of singular gauge requires placing the theory on punctured Euclidean spacetime, which whilst flat, nonetheless also differs on a topological level from unpunctured spacetime \cite{Dorey:2002ik}.
Needless to say, this does not however impede the application of standard methods.}.
Since the topological invariants of supergravity are exactly those of general relativity \cite{Townsend:1979js}, albeit allowing for torsion, from \eqref{topological invariants} we then find
\begin{align}
	\chi&=\frac{1}{16\pi^2}\int_\mathcal{M} \,R_{ab}\wedge *R^{ab}
	=\frac{1}{4\pi^2}\int_\mathcal{M} F^i\wedge *F^{i}
	=|k| \,,\nonumber\\
	\label{fermionic invariants}
	\tau
	&=\frac{1}{24\pi^2}\int_\mathcal{M} \,R_{ab}\wedge R^{ab}
	=\frac{1}{6\pi^2}\int_\mathcal{M} \,F^i\wedge F^{i}
	=\frac{2}{3}k\,,
\end{align}
where $k$ is the instanton number from \eqref{Pontryagin number}
\footnote{\,That $\chi=\frac{3}{2}|\tau|$ here is coherent with the general properties of torsionful Riemannian geometry \cite{Ferreira:2011zzb}.}
\footnote{\,Our $\mathfrak{su\left(2\right)}$ convention is $T^a=i\vec\tau$, such that $\mathrm{Tr}\left(T^aT^b\right)=-\delta^{ab}$ and $\left[T^a,T^b\right]=-2\epsilon^{abc}T^c$.}.

Since we may construct every self-dual Yang-Mills connection here via the ADHM method, we may therefore construct all corresponding fermionic instantons on this background.
Reinstating the gravitational coupling
\footnote{\,The dimensionally corrected supersymmetry variation $\delta\psi=D\epsilon/\kappa$ implies that $[\epsilon]=-1/2$.},
this yields in the simplest case
\begin{align}
	\psi_+
	=\frac{1}{\kappa}A\epsilon_+
	=\frac{1}{\kappa}\frac{x^a\sigma_{ab}}{x^2+\rho^2}\,e^b\,\epsilon_+ \,,
	\label{BPST gravitino}
\end{align}	
where $\epsilon_+$ is a constant spinor, and the analogy with \eqref{BPST instanton} is clear.

We may summarise a number of important further properties.
\begin{enumerate}
	\item By virtue of the construction $\psi_+$ must have two degrees of freedom, exactly those of a massless gauge field and importantly, the requisite number for one half of a Dirac vector-spinor.
	As any such gauge field is transverse, $\psi_+$ furthermore is.
	\item Given their straightforward proportionality, the moduli space of these fermionic instantons is inherited from the corresponding Yang-Mills equivalent.
	In the case of \eqref{BPST gravitino} we have eight moduli; four coordinates denoting the centre of the instanton, three determining an orientation within $SU\left(2\right)_+$, and one scale size.
	Generalising to higher instanton numbers yields in general $8k$ moduli.
	\item Compatibility with the Atiyah-Singer index theorem is self evident, in that for some arbitrary spinor $D\chi_+=d\chi_++A\wedge\chi_+$, exactly as we would have for a spinor in flat torsion-free spacetime in the background of a Yang-Mills instanton.
	The number of zero modes of the Dirac operator is then correlated with the $8k$ collective coordinates $A$ possesses.
	\item Since $A$ is asymptotically pure gauge, $\epsilon_+$ is asymptotically Killing, and $\psi_+$ is accordingly asymptotic to the supersymmetry transformation $\delta\psi=D\epsilon_+=0$ at infinity.
	\item The global $U(1)$ chiral symmetry of $\mathcal{N}=2$ supergravity is violated by these solutions, as expected from the corresponding anomalous divergence relation $D*J\sim R^{ab}\wedge *R_{ab}$.
	\end{enumerate}

Although for reasons of clarity we have specialised latterly to the case of zero spacetime curvature, the underlying construction presented in the previous sections is of course valid for general antiself-dual backgrounds. 
Existing, and in some cases exhaustive, constructions of Yang-Mills instantons on ALE/ALF spaces \cite{NahmCalorons,KN,Bianchi:1995xd, Kraan:1998xe,Cherkis:2009jm,Cherkis:2010bn} may then be leveraged to arrive at corresponding fermionic solutions. 
Conceivably one may also implement the extension of the ADHM method to non-commutative $\mathbb{R}^4$ \cite{Nekrasov:1998ss} to provide analogous solutions in non-commutative generalisations of supergravity \cite{Aschieri:2009mc} 
\footnote{\,Intriguingly the requirement of a chiral gravitino, which motivates the present setting of $\mathcal{N}=2$ supergravity, is in fact enforced by the non-commutative $\mathcal{N}=1$ local supersymmetry in four dimensions.}.

In more general backgrounds, the topological invariants in \eqref{fermionic invariants} will of course receive contributions from both bosonic and fermionic sectors.
Since the curvature decomposition in \eqref{Riemann decomposition} is orthogonal, by virtue of mutually incompatible duality conditions, we have
\begin{align}
	\chi=\chi_\circ+|k|\,,\quad
	\tau=\tau_\circ+\frac{2}{3}k\,,
\end{align}
where $\chi_\circ$ and $\tau_\circ$ are respectively the Euler number and Hirzebruch signature of the gravitational background in question.

\section{Conclusions}

We have demonstrated the existence of a broad class of zero action, half-BPS solutions to the Euclidean equations of motion of supergravity, which we identify as fermionic instantons associated to the status of the gravitino as the gauge field of local supersymmetry.

These may be explicitly constructed in practice from self-dual Yang-Mills gauge fields and the Killing spinors of the torsion-free background.
By leveraging existing results from instanton calculus in this way we may construct all possible solutions on certain backgrounds, including asymptotically locally Euclidean spaces.
A further possibility also exists to implement the construction of instantons in non-commutative $\mathbb{R}^4$ within the known non-commutative generalisations of supergravity.

These fermionic instantons thereby constitute a novel class of topologically non-trivial supergravity backgrounds, which in contrast to conventional `bosonic' solutions feature non-vanishing fermionic character, and as a necessary consequence, spacetime torsion.

As an interesting aside, such backgrounds provide an explicit realisation of the scenario explored in \cite{Hanson:1978eg, D'Auria:1981yg} seeking to relate quantum gravity and superconductivity via spacetime torsion. 
Therein, a gravitational Meissner effect was invoked to generate microscopic torsion vortices in an analogous fashion to the creation of magnetic flux vortices in type II superconductors, albeit at the cost of introducing a number of somewhat contrived features, including a spacetime dependent cosmological constant. 
In the present context these torsion vortices are supplied by the gravitino instanton cores, which are in contrast natural phenomena within the context of supergravity.

Given the wide range of existing instanton applications and literature, and the fact that the solutions presented herein inherit their form from `conventional' gauge theory instantons, the primary task at hand seems to be to now import known results into this novel fermionic context.
This logic may of course also be extended beyond instantons to other topologically non-trivial phenomena, such as monopoles, vortices and domain walls.
Of course one needn't limit attention to topologically special solutions; we may also leverage these considerations to search for fermionic configurations of more prosaic character.

Going beyond the present minimal setting, we also note that there exists a wide array of supergravity and string theories in diverse dimensions, possessing varied field content and symmetry.
The utility and behaviour of these new solutions within such contexts is also naturally of interest.

One further aspect of particular importance is the need to understand the role of these solutions in the super-Higgs phase, given the relevance that this may have for modern-day phenomenology.

In closing, we also note that it is not necessarily guaranteed that the construction given here of fermionic instantons is complete.
Conceivably there may also be solutions which are not globally of the form $\psi=D\epsilon$.
The existence and possible properties of such configurations is then also an important open question.

\section*{Acknowledgements}

We wish to thank Richard Szabo and Tianjun Li for useful comments and proofreading.


\begin{thebibliography}{99}

\bibitem{Belavin:1975fg}
  A.~A.~Belavin, A.~M.~Polyakov, A.~S.~Schwartz and Y.~S.~Tyupkin,
  ``Pseudoparticle Solutions of the Yang-Mills Equations,''
  Phys.\ Lett.\ B {\bf 59} (1975) 85.

\bibitem{Schafer:1996wv}
  T.~Sch\"afer and E.~V.~Shuryak,
  ``Instantons in QCD,''
  Rev.\ Mod.\ Phys.\  {\bf 70} (1998) 323
  [hep-ph/9610451].
  
\bibitem{Bianchi:2007ft}
  M.~Bianchi, S.~Kovacs and G.~Rossi,
  ``Instantons and Supersymmetry,''
  Lect.\ Notes Phys.\  {\bf 737} (2008) 303
  [hep-th/0703142].
  
\bibitem{Freed:1984xe}
  D.~S.~Freed and K.~K.~Uhlenbeck,
  ``Instantons And Four - Manifolds,''
  New York, Usa: Springer ( 1984) 232 P. ( Mathematical Sciences Research Institute Publications, 1)
  
\bibitem{Eguchi:1978gw}
  T.~Eguchi and A.~J.~Hanson,
  ``Selfdual Solutions to Euclidean Gravity,''
  Annals Phys.\  {\bf 120} (1979) 82.
 
\bibitem{Gibbons:1982fy}
  G.~W.~Gibbons and C.~M.~Hull,
  ``A Bogomolny Bound for General Relativity and Solitons in N=2 Supergravity,''
  Phys.\ Lett.\ B {\bf 109} (1982) 190.
  
\bibitem{Tod:1983pm}
  K.~p.~Tod,
  ``All Metrics Admitting Supercovariantly Constant Spinors,''
  Phys.\ Lett.\ B {\bf 121} (1983) 241.
    
\bibitem{Gauntlett:2002nw}
  J.~P.~Gauntlett, J.~B.~Gutowski, C.~M.~Hull, S.~Pakis and H.~S.~Reall,
  ``All supersymmetric solutions of minimal supergravity in five- dimensions,''
  Class.\ Quant.\ Grav.\  {\bf 20} (2003) 4587
  [hep-th/0209114].
  
\bibitem{FigueroaO'Farrill:2002ft}
  J.~M.~Figueroa-O'Farrill and G.~Papadopoulos,
  ``Maximally supersymmetric solutions of ten-dimensional and eleven-dimensional supergravities,''
  JHEP {\bf 0303} (2003) 048
  [hep-th/0211089].
  
\bibitem{Gauntlett:2002fz}
  J.~P.~Gauntlett and S.~Pakis,
  ``The Geometry of D = 11 killing spinors,''
  JHEP {\bf 0304} (2003) 039
  [hep-th/0212008].
  
\bibitem{Gutowski:2003rg}
  J.~B.~Gutowski, D.~Martelli and H.~S.~Reall,
  ``All Supersymmetric solutions of minimal supergravity in six- dimensions,''
  Class.\ Quant.\ Grav.\  {\bf 20} (2003) 5049
  [hep-th/0306235].
    
\bibitem{Bellorin:2005zc}
  J.~Bellorin and T.~Ortin,
  ``All the supersymmetric configurations of N=4, d=4 supergravity,''
  Nucl.\ Phys.\ B {\bf 726} (2005) 171
  [hep-th/0506056].

\bibitem{Cacciatori:2007vn}
  S.~L.~Cacciatori, M.~M.~Caldarelli, D.~Klemm, D.~S.~Mansi and D.~Roest,
  ``Geometry of four-dimensional Killing spinors,''
  JHEP {\bf 0707} (2007) 046
  [arXiv:0704.0247].
    
\bibitem{Bellorin:2007yp}
  J.~Bellorin and T.~Ortin,
  ``Characterization of all the supersymmetric solutions of gauged N=1, d=5 supergravity,''
  JHEP {\bf 0708} (2007) 096
  [arXiv:0705.2567].
 
 \bibitem{Gran:2008vx}
  U.~Gran, J.~Gutowski and G.~Papadopoulos,
  ``Geometry of all supersymmetric four-dimensional N = 1 supergravity backgrounds,''
  JHEP {\bf 0806} (2008) 102
  [arXiv:0802.1779].

\bibitem{Urrutia:1981sk}
  L.~F.~Urrutia,
  ``A New Exact Solution of Classical Supergravity,''
  Phys.\ Lett.\ B {\bf 102} (1981) 393.
 
\bibitem{Hull:1983aw}
  C.~M.~Hull,
  ``Killing spinors and exact plane wave solutions of extended supergravity,''
  Phys.\ Rev.\ D {\bf 30} (1984) 334.
    
\bibitem{Hull:1984vh}
  C.~M.~Hull,
  ``Exact $p p$ Wave Solutions of Eleven-dimensional Supergravity,''
  Phys.\ Lett.\ B {\bf 139} (1984) 39.
   
\bibitem{Aichelburg:1986wv}
  P.~C.~Aichelburg and F.~Embacher,
  ``The Exact Superpartners of $N=2$ Supergravity Solitons,''
  Phys.\ Rev.\ D {\bf 34} (1986) 3006.
 
\bibitem{Dorey:2002ik}
  N.~Dorey, T.~J.~Hollowood, V.~V.~Khoze and M.~P.~Mattis,
  ``The Calculus of many instantons,''
  Phys.\ Rept.\  {\bf 371} (2002) 231,
  [hep-th/0206063].
  
\bibitem{Osterwalder:1973dx}
  K.~Osterwalder and R.~Schrader,
  ``Axioms For Euclidean Green's Functions,''
  Commun.\ Math.\ Phys.\  {\bf 31} (1973) 83;\\
    K.~Osterwalder and R.~Schrader,
  ``Axioms for Euclidean Green's Functions. 2.,''
  Commun.\ Math.\ Phys.\  {\bf 42} (1975) 281.

\bibitem{Zumino:1977yh}
  B.~Zumino,
  ``Euclidean Supersymmetry and the Many-Instanton Problem,''
  Phys.\ Lett.\ B {\bf 69} (1977) 369.
    
\bibitem{Nicolai:1978vc}
  H.~Nicolai,
  ``A Possible constructive approach to (Super $\phi^3$) in four-dimensions. 1. Euclidean formulation of the model,''
  Nucl.\ Phys.\ B {\bf 140} (1978) 294.
  
\bibitem{Mehta:1986mi}
  M.~R.~Mehta,
  ``Euclidean Continuation of the Dirac Fermion,''
  Phys.\ Rev.\ Lett.\  {\bf 65} (1990) 1983
   Erratum: [Phys.\ Rev.\ Lett.\  {\bf 66} (1991) 522].
    
\bibitem{vanNieuwenhuizen:1996tv}
  P.~van Nieuwenhuizen and A.~Waldron,
  ``On Euclidean spinors and Wick rotations,''
  Phys.\ Lett.\ B {\bf 389} (1996) 29
  [hep-th/9608174].
    
\bibitem{Fubini:1972mf}
  S.~Fubini, A.~J.~Hanson and R.~Jackiw,
  ``New approach to field theory,''
  Phys.\ Rev.\ D {\bf 7} (1973) 1732.
    
\bibitem{Deser:1977ur}
  S.~Deser, J.~H.~Kay and K.~S.~Stelle,
  ``Hamiltonian Formulation of Supergravity,''
  Phys.\ Rev.\ D {\bf 16} (1977) 2448.
  
\bibitem{Witten:1981nf}
  E.~Witten,
  ``Dynamical Breaking of Supersymmetry,''
  Nucl.\ Phys.\ B {\bf 188} (1981) 513.
  
\bibitem{Charap:1977ww}
  J.~M.~Charap and M.~J.~Duff,
  ``Gravitational Effects on Yang-Mills Topology,''
  Phys.\ Lett.\ B {\bf 69} (1977) 445.
   
\bibitem{Atiyah:1978ri}
  M.~F.~Atiyah, N.~J.~Hitchin, V.~G.~Drinfeld and Y.~I.~Manin,
  ``Construction of Instantons,''
  Phys.\ Lett.\ A {\bf 65} (1978) 185.
    
  \bibitem{Freedman:2012zz}
  D.~Z.~Freedman and A.~Van Proeyen,
  ``Supergravity,''
  Cambridge, UK: Cambridge Univ. Pr. (2012).
  
  \bibitem{Atiyah:1978wi}
  M.~F.~Atiyah, N.~J.~Hitchin and I.~M.~Singer,
  ``Selfduality in Four-Dimensional Riemannian Geometry,''
  Proc.\ Roy.\ Soc.\ Lond.\ A {\bf 362} (1978) 425.
  
\bibitem{Townsend:1979js}
  P.~K.~Townsend and P.~van Nieuwenhuizen,
  ``Anomalies, Topological Invariants and the {Gauss-Bonnet} Theorem in Supergravity,''
  Phys.\ Rev.\ D {\bf 19} (1979) 3592.
   
\bibitem{Ferreira:2011zzb}
  A.~C.~Ferreira,
  ``Einstein four-manifolds with skew torsion,''
  J.\ Geom.\ Phys.\  {\bf 61} (2011) 2341.
  
  \bibitem{NahmCalorons}   
  W.~Nahm,
  ``Selfdual Monopoles and Calorons,''
BONN-HE-83-16
{\it Presented at 12th Colloq. on Group Theoretical Methods in Physics, Trieste, Italy, Sep 5-10, 1983};\\
W. Nahm, ``Self-dual Monopoles and Calorons,''
Physics {\bf 201}, Springer, New York, 1984, pp. 189--200.
  
  \bibitem{KN}
P.~B.~Kronheimer and H.~Nakajima ``Yang-Mills Instantons on ALE Gravitational
Instantons,'' Math. Ann. {\bf 288} (1990), no. 2, 263--307.

\bibitem{Bianchi:1995xd}
  M.~Bianchi, F.~Fucito, G.~Rossi and M.~Martellini,
  ``On The ADHM Construction on ALE Gravitational Backgrounds,''
  Phys.\ Lett.\  B {\bf 359}, 49 (1995);

  M.~Bianchi, F.~Fucito, G.~Rossi and M.~Martellini,
  ``Explicit Construction of Yang-Mills Instantons on ALE Spaces,''
  Nucl.\ Phys.\  B {\bf 473}, 367 (1996)
  [hep-th/9601162].
  
    \bibitem{Kraan:1998xe}
  T.~C.~Kraan and P.~van Baal,
  ``New Instanton Solutions at Finite Temperature,''
  Nucl.\ Phys.\  A {\bf 642}, 299 (1998)
  [hep-th/9805201];
\\
  T.~C.~Kraan and P.~van Baal,
  ``Periodic Instantons with Non-trivial Holonomy,''
  Nucl.\ Phys.\  B {\bf 533}, 627 (1998)
  [hep-th/9805168];
\\
  T.~C.~Kraan and P.~van Baal,
  ``Exact T-duality between Calorons and Taub - NUT Spaces,''
  Phys.\ Lett.\  B {\bf 428}, 268 (1998)
  [hep-th/9802049].
  
   \bibitem{Cherkis:2009jm}
  S.~A.~Cherkis,
  ``Instantons on the Taub-NUT Space,''
  Adv.\ Theor.\ Math.\ Phys.\  {\bf 14} (2010) no.2,  609
  [arXiv:0902.4724].
    
  \bibitem{Cherkis:2010bn}
  S.~A.~Cherkis,
  ``Instantons on Gravitons,''
  Commun.\ Math.\ Phys.\  {\bf 306} (2011) 449
  [arXiv:1007.0044].
       
\bibitem{Nekrasov:1998ss}
  N.~Nekrasov and A.~S.~Schwarz,
  ``Instantons on noncommutative $\mathbb{R}^4$ and (2,0) superconformal six-dimensional theory,''
  Commun.\ Math.\ Phys.\  {\bf 198} (1998) 689
  [hep-th/9802068].
    
\bibitem{Aschieri:2009mc}
  P.~Aschieri and L.~Castellani,
  ``Noncommutative supergravity in D=3 and D=4,''
  JHEP {\bf 0906} (2009) 087
  [arXiv:0902.3823].
    
\bibitem{Hanson:1978eg}
  A.~J.~Hanson and T.~Regge,
  ``Torsion And Quantum Gravity,''
  LBL-8240.
  
\bibitem{D'Auria:1981yg}
  R.~D'Auria and T.~Regge,
  ``Gravity Theories With Asymptotically Flat Instantons,''
  Nucl.\ Phys.\ B {\bf 195} (1982) 308.
    
\end{thebibliography}
\end{document}